*Title*: Selective activation of resting state networks following focal stimulation in a connectome-based network model of the human brain

*Abbreviated title*: Stimulation in connectome-based brain models

*Author names and affiliation*:
[1] Andreas Spiegler
   Aix Marseille Université, Institut de la Santé et de la Recherche Médical, Institut de Neurosciences des Systèmes UMR_S 1106, 13005, Marseille, France
[2] Enrique C.A. Hansen
   Aix Marseille Université, Institut de la Santé et de la Recherche Médical, Institut de Neurosciences des Systèmes UMR_S 1106, 13005, Marseille, France
[3] Christophe Bernard
   Aix Marseille Université, Institut de la Santé et de la Recherche Médical, Institut de Neurosciences des Systèmes UMR_S 1106, 13005, Marseille, France
[4] Anthony R. McIntosh
   Rotman Research Institute of Baycrest Center, University of Toronto, Toronto, M6A 2E1, Canada
[5] Viktor K. Jirsa
   Aix Marseille Université, Institut de la Santé et de la Recherche Médical, Institut de Neurosciences des Systèmes UMR_S 1106, 13005, Marseille, France

*Corresponding authors*:
Andreas Spiegler
   Aix Marseille Université, Institut de la Santé et de la Recherche Médical, Institut de Neurosciences des Systèmes UMR_S 1106, 13005, Marseille, France
   andreas.spiegler@univ-amu.fr
Viktor Jirsa
   Aix Marseille Université, Institut de la Santé et de la Recherche Médical, Institut de Neurosciences des Systèmes UMR_S 1106, 13005, Marseille, France
   viktor.jirsa@univ-amu.fr

*Number of pages*: 23

*Number of figures*: 7

*Number of tables*: 2

*Number of words for abstract*: 201

*Number of words for significance statement*: 88

*Number of words for introduction*: 646

*Number of words for discussion*: 994

*Total number of words*: 7859

*Conflict of interest*: The authors declare no competing financial interests.

*Acknowledgements*: The research reported herein was supported by the Brain Network Recovery Group through the James S. McDonnell Foundation and funding from the European Union Seventh Framework Programme: FP7-ICT Human Brain Project (grant no. 60 402).

**Abstract**

Imaging studies suggest that the functional connectivity patterns of resting state networks (RS-networks) reflect underlying structural connectivity (SC). If the connectome constrains how brain areas are functionally connected, the stimulation of specific brain areas should produce a characteristic wave of activity ultimately resolving into RS-networks. To systematically test this hypothesis, we use a connectome-based network model of the human brain with detailed realistic SC. We systematically activate all possible thalamic and cortical areas with focal stimulation patterns and confirm that the stimulation of specific areas evokes network patterns that closely resemble RS-networks. For some sites, one or no RS-network is engaged, whereas for other sites more than one RS-network may evolve. Our results confirm that the brain is operating at the edge of criticality, wherein stimulation produces a cascade of functional network recruitments, collapsing onto a smaller subspace that is constrained in part by the anatomical local and long-range SCs. We suggest that information flow, and subsequent cognitive processing, follows specific routes imposed by connectome features, and that these routes explain the emergence of RS-networks. Since brain stimulation can be used to diagnose/treat neurological disorders, we provide a look-up table showing which areas need to be stimulated to activate specific RS-networks.

**Significance statement**

Systematic exploration via stimulation of all cortical and subcortical brain areas can only be performed *in silico*. We have performed a detailed parametric exploration of a large-scale brain network model and developed a stimulation map indicating which brain areas need to be stimulated to place the brain in a particular state at rest. Brain stimulation is one of the upcoming novel tools in the treatment of neurological disorders. The stimulation map will be critical in guiding these studies and allow for the development of theory guided stimulation protocols.

## Introduction

Local and global computation in the brain strongly depends upon local (intracortical) and long-range (intercortical) structural connections (aka the large-scale connectome). Intuitively, the connectome imposes strong constraints upon where and how information will be transmitted and distributed (Deco et al., 2015), but direct evidence is still lacking. Imaging studies provide support for this hypothesis. In the absence of external inputs or goal-oriented tasks, that is, during rest, several brain areas are functionally connected to each other, thus defining resting state networks (RS-networks). These functional networks correlate with the structural connectivity (SC) of white matter tracts (van den Heuvel et al., 2009; Greicius et al., 2009; Hermundstad et al., 2013). It thus appears that, RS-networks merely reflect the basic topological features of the connectome. If this hypothesis is correct, we predict that the activation of a given brain area will give rise to a wave of activity ultimately resolving in RS-networks. Testing this hypothesis experimentally is difficult as it requires knowing where and how to stimulate. As a first step in this direction, we choose an unbiased computational approach, using a large-scale brain network model to stimulate every part of the brain and map the dissipation of the stimulation to the RS-networks.

Brain networks have specific constraints due to the large-scale spatiotemporal scale of operation. Firstly, the time delays due to signal transmission via long white matter tracts between connecting nodes in brain network dynamics have been suggested to play a crucial role in the generation of RS-activity (Ghosh et al., 2008). Hence, without veridical time delays, the dynamics may not resemble empirical RS-networks. Secondly, the connection strength, when scaled appropriately, places the brain close to criticality where the capacity of processing information is maximized and the functional connectivity (FC) has been found to best fit to empirical data at rest (Deco et al., 2014a; Deco and Jirsa, 2012; Ghosh et al., 2008). Finally, random processes serve to provide the brain with kinetic energy to form and alter functional networks (Deco et al. 2014a, Ghosh et al., 2008). Previous studies mostly considered long-range connections (i.e., SC of white matter tracts). We go beyond this and incorporate local SC to understand how activity propagates and dissipates in the brain (Qubbaj and Jirsa, 2007, 2009; Jirsa, 2004; Jirsa and Kelso, 2000). We use the simulation platform *The Virtual Brain* (TVB) to study areal dynamics through mean field approximations (Sanz-Leon et al., 2013, 2015), following focal stimulation, taking into account both long-range and local SCs. Because certain structural parameters are not known quantitatively, we parametrically explore all possibilities, in particular the ratio of long-range to local SCs, and the spatial range of local SC (Spiegler and Jirsa, 2013).

To study RS-dynamics, brain areas need to operate near criticality (Ghosh et al., 2008; Deco et al., 2011, 2013). Near-criticality is defined as a system that is on the brink of a qualitative change in its behavior (Shew and Plenz, 2013). The proximity to criticality predicts that the brain's response to stimulation will primarily arise from structures and networks that are closest to instability. Activities in those networks require the most time to settle into equilibria after stimulation, and are associated with large-scale dependencies and scale invariance (Haken, 1978). This would be consistent with the center manifold theorem, which states that a high-dimensional system in a subcritical state will converge on a lower dimensional manifold (here few networks) when the system is stimulated.

Using an unbiased approach, here we demonstrate that evoked network activity dissipates into experimentally known RS-networks (Damoiseaux et al., 2006) in a non-trivial manner. In particular, we show that stimulation at spatially distant sites can give rise to similar non-stationary trajectories, whereas stimulation at spatially close sites can result in distinctly different dynamics. We provide a look-up table to target specific areas to activate specific networks, an important information for future clinical trials using brain stimulations to restore function.

**Materials and methods**

Using *The Virtual Brain* platform (Sanz-Leon et al., 2013, 2015), we triangulate the surface of the cortex with a mesh of 8,192 nodes for each hemisphere (**Fig. 1a**), distributed across 74 cortical areas (**Fig. 1b**), each containing between 29 and 683 nodes (Table 1), following a known functional parcellation atlas (Kötter and Wanke, 2005). The model also includes non-parcellated 116 subcortical areas. To connect nodes with each other, we distinguish homogeneous from heterogeneous SC (**Fig. 1c–e**). The homogenous SC (of short-range connections) links nodes within an area, and between areas if they are spatially close from one another with a connection probability decreasing with distance (Braitenberg and Schüz, 1991) (**Fig. 1c**, and **d**). The heterogeneous SC (of long-range white matter tracts) links all the nodes of an area with the nodes of another area (**Fig. 1c and e**), based on known anatomical data (Kötter and Wanke, 2005). Although neighboring areas may not be connected via the white matter tract (i.e., heterogeneous SC), they are still able to exchange information via the homogeneous SC (e.g. Area 2 with Areas 1 and 3 in **Fig. 1c**).

Each vertex point is a network node holding a neural mass model connected to other nodes via the homogeneous and heterogeneous SCs. When an area is stimulated, all the nodes of this area are simultaneously activated and then the stimulation-induced activity in each node decays differently according to the activity in the surrounding via short-range connections (i.e., homogeneous SC) and remote nodes via long-range connections (i.e., heterogeneous SC). Because the ratio of homogeneous to heterogeneous SCs is not known, we consider this ratio as a degree of freedom and perform a parametric study (see Jirsa and Kelso, 2000; Qubbaj and Jirsa, 2007, 2009 for systematic studies with two-point connection). At the extremes, (i) 0 % of heterogeneous SC (thus 100 % of homogeneous SC gives two unconnected cerebral hemispheres with locally but homogeneously connected nodes) only allows activity to propagate locally from a cortical stimulation site, and (ii) 100 % of heterogeneous SC (thus 0 % of homogenous SC gives 190 purely heterogeneously connected brain areas with locally unconnected nodes) only allows activity to travel long distances with time delays via white matter fiber tracts. Furthermore, since the spatial range of local connections, that is, the homogeneous SC is not known, we also consider it as a parameter varying between 10 mm and 41 mm (Spiegler and Jirsa, 2013). We then systematically stimulate each of the 190 areas with a large range of parameter values (for the ratio and the spatial range), resulting in a total of all 37,620 simulation trials.

We use the near-criticality regimen to trigger, via stimulation, a perturbation of a given area in the direction of its instability point and induce characteristic energy dissipation through the brain network. The dissipation of energy will be constrained by the homogeneous and heterogeneous SCs, the associated signal transmission delays, and the local dynamics at the network nodes. In the network model, the operating point of every node, when disconnected from the network, is at the same distance from the critical point (**Fig. 2a**). If the critical point is reached, the node enters into a constant oscillatory mode. In the network, the SCs (incl. time delays) determine the alteration of the working distance to the critical point at each node in time by weighting the delaying the incoming activity from other nodes in the network. The network model, however, is set so that criticality is never reached. As a result, when a node is stimulated, the node operates closer to the critical point and the response is in the form of a damped oscillation (**Fig. 2a**). The closer a node operates to the critical point, the stronger the node's responses with high amplitude and long decay time (**Fig. 2a**). The nodes are working near criticality (i.e., they get close to a change in behavior, which would be here a switch to a constant oscillatory mode, but never reaching it). Thus the response to the stimulation is transient, lasting a few milliseconds. The damped oscillation generated in one stimulated node is then sent via its efferent connections to its target nodes, triggering there, in turn, a damped oscillation (**Fig. 2b**). If the network were mainly based on nodes connected in series, activity would decay very fast after the stimulation (**Fig. 2b**). However, since the outgoing activity of a node can influence the nodes projecting back to it, recurrent systems appear (**Fig. 2c, d**), which allow activity to dissipate on a much longer time scale. The evoked activity, after the initial decay, thus persists in the so-called dissipation networks (**Fig. 2c, d**), which may reflect feedback loops and re-entry points in the SC. This dissipation property is shown on **Fig. 3a**. The activation of three different areas gives rise to three different responses in a given target area. The differences stem from the proximity to criticality, which depends upon the SCs (in particular the extent of recurrent networks), comprising the synaptic weights and the time delays (**Fig. 1**). This behavior is predicted by the center manifold theorem, which is the mathematical basis for criticality (Haken, 1978).

*Large-scale brain model.* Dynamics of a vector field $\Psi(x, t)$ at time $t \in \mathbb{R}^1$ and position $x \in \mathbb{R}^3$ in space $\Omega$ are described by a delay-integro-differential equation:

$$\begin{aligned}\partial_t \Psi(x,t) = &\; E\big(\Psi(x,t)\big) - a_I I(x,t) \\ &+ (1-\alpha)\, a_L \int_L dX'\, \Psi(x-X', t)\, g(X') \\ &+ \alpha\, a_S \int_\Omega dX'\, \Psi(x-X', t - \|x-X'\|/v) \\ &\quad \times\; H(x)\, C(\|x-X'\|/v)\, K^T(X') \end{aligned} \quad , \qquad (1)$$

were $\partial_t$ is the derivative with respect to time, $t$. The input $I(x, t)$ allows the stimulation dynamics to intervene on a node. The operator $E(\Psi(x, t))$ locally links variables of the vector field. The scalar $\alpha$ balances the effect of the homogeneous and the heterogeneous SCs (first and second integral) on the vector field. The vectors $a_I$, $a_L$, and $a_S$ of factors relate the input $I$, and both SCs to the vector field $\Psi(x, t)$. The kernel $g(x)$ describes the homogeneous SC. The field is time delayed due to a finite transmission speed $v$ via the heterogeneous SC given by matrix $C(x)$. The vectors $H(x)$ and $K(x)$ establish the links between the heterogeneous SC and the targets and the sources. The spatial and temporal aspects of the model are described in more detail in the following two subsections.

*Geometry and structural connectivity (SC).* The spatial domain $\Omega = \{L_1 \cup L_2 \cup S\}$ separates both cerebral hemispheres $L = \{L_1 \cup L_2\}$: left, $L_1$ and right, $L_2$, from subcortical areas $S$, that is, $\cap \Omega = \varnothing$. A closed 2-sphere describes the geometry of each hemisphere ($L_1$ and $L_2$). The homogeneous

SC follows a Gaussian distribution $g(x) = \exp(-x^2/(2\sigma^2))$ invariant under translations on $L$ (Spiegler and Jirsa, 2013). Each closed sphere, $L_1$ and $L_2$, is divided into $m = 38$ regions, $L_1 = \cup_{r \in R_1} A_r$ and $L_2 = \cup_{r \in R_2} A_r$ with $R_1 = R(m)$, $R_2 = R_1 + n : R(\lambda \in \mathbb{N}) = \{r \mid r \in \mathbb{N}, r \leq \lambda\}$ following a coarser Brodmann map (Kötter and Wanke, 2005) of areas, $A_r = A(r \in \mathbb{N}) \in \Omega : \mathbb{N} \to \mathbb{R}^3$ onto space $\Omega$ for introducing heterogeneous SC. The corpus callosum intersects the medial faces of both closed 2-spheres to interconnect both cerebral hemispheres from within, leaving two openings. The region on each sphere where the corpus callosum intersects is simply taken out by placing all nodes in these two regions far enough apart so that the homogeneous SC of these nodes is $g(x - X') \to 0$. Finally, one region is the intersection by the corpus callosum and the remaining regions are the considered 37 cortical areas composing a cerebral hemisphere. Each of the $n = 116$ considered subcortical areas is lumped to a single point in space $S = \cup_{r \in R_3} A_r$ with $R_3 = R(n) + 2m$. The heterogeneous connections, $C$ transmit mean activities of sources to target areas, $H(x)$ and $K(X')$ with a finite transmission speed, $v = 6$ ms$^{-1}$ (Nunez, 1995, 1981). The square matrix, $C(\|x - X'\|/v)$ contains $(2m + n)^2$ weights, $c_{ij}(\|x - X'\|/v) : i, j = 1, \ldots, 2m + n$ taken from the *CoCoMac* database (Kötter et al., 2004, 2005; Stephan et al., 2001) which was extrapolated to human. The row vectors $H(x)$ and $K(X')$ contain $2m + n$ operations, $h_i(x)$ and $k_j(X')$ on the targets and sources, respectively. The operations are $h_i(x) = \delta_x(A_i)$ and $k_j(X') = \delta_{X'}(A_j)/|A_j|$ with the Dirac measure $\delta_\Omega(A)$ on $\Omega$ and the cardinality $|A_r|$ of the set $A_r$.

*Temporal dynamics*. The vector field describes a two-dimensional flow (Stefanescu and Jirsa, 2008) linking two variables $\Psi(x, t) = (\psi_1 \ \psi_2)^T(x, t)$ in (1) as follows

$$E(\Psi(x,t)) = \eta \begin{pmatrix} \psi_2(x,t) - \gamma\, \psi_1(x,t) - \psi_1^3(x,t) \\ -\varepsilon\, \psi_1(x,t) \end{pmatrix}, \tag{2}$$

The parameterization: $\gamma = 1.21$ and $\varepsilon = 12.3083$ sets an isolated brain area close to a critical point, that is, an Andronov-Hopf bifurcation (sketched in **Fig. 2**) with a natural frequency around 42 Hz using a characteristic rate of $\eta = 76.74$ s$^{-1}$. This rhythm in the gamma band accounts for local activity such as a coordinated interaction of excitation and inhibition (Buzsáki and Wang, 2012) that is not explicitly modeled here. The Dirac delta function is applied to a brain area, $I_r(x, t) = -5\eta\, \delta_x(A_r)\, \delta(t)$. The connectivities and the input act on the first variable $\psi_1(x, t)$ in (1) by $a_L = a_S = (a_I)^T = (\eta \ 0)$. The connectivity-weighted input determines criticality by working against inherent energy dissipation (i.e., stable focus) towards the bifurcation. So that the bifurcation was not passed, both homogeneous and heterogeneous SCs, $g(x)$ and $C(\|x - X'\|/v)$ are normalized to unity maximum in-strength across time delays by: (i) $\int dx\, g(x) = 1$, and (ii) $\sup_{\lambda \in \Omega/v} \{\sum_j^n c_{ij}(\|\lambda\|)\} = 1$.

*Simulation.* To simulate the model on a digital computer, physical space and time are discretized. The folding of the human cortex presents a challenge for sampling. The cerebral surfaces, $L_1$ and $L_2$, are evenly filled with 8,192 nodes. Subcortical structures in $S$ remain unaffected by the discretization. The geometry of the brain is captured in physical space, $\Omega$ by a net of 16,500 nodes (i.e. 16,384 cortical and 116 subcortical). The spatial integrals in (1) are rewritten as matrix operations, where the heterogeneous SC remains the same and the homogeneous SC is spatially sampled on the cerebral surfaces (Spiegler and Jirsa, 2013). The system of difference equations are then solved using Heun's method with a time step of 40 μs for 1 second per realization of one of the following factors: each of the 190 stimulation sites, SC-balance, $\alpha = \{0.0, 0.2, 0.4, 0.6, 0.8, 1.0\}$, and homogeneous spreading, $\sigma / mm \in \mathbb{N} : 10 \leq \sigma / mm \leq 41$. The implementation is verified by the algebraic solution of an isolated node (i.e., no connections), and by the field properties (e.g., compact solutions spreading radially around a stimulation site) of the homogeneously linked cerebral nodes.

*Cellular automaton.* The transient period after stimulation onset caused by the transmission times among the 190 brain areas (74 cortical and 116 subcortical areas) in the heterogeneous SC is estimated using a cellular automaton. Since the transmissions are assumed to be instantaneous in each cerebral hemisphere among the 8,192 nodes through the homogeneous SC, the cerebral areas and its transmission times are considered to cause transients rather than each single node in cerebral areas and its short-range interactions. Each of the 190 cells in the cellular automaton describes one of the brain area given by the homogeneous SC to be either active or inactive. The temporal decomposition of the heterogeneous SC according to the transmission times gives rules for changing the state of cells over time. The cellular automaton is initialized from the overall inactive state. An activation of a cell triggers a cascade of activation in time until no more cells get activated. In this manner 190 characteristic activation cascades emerged, each by stimulation, that is, activation of a single cell. The time that the cellular automaton enters the steady state across all stimulation estimates the transient period from the time delays in the heterogeneous SC.

*Stimulation and decomposition.* All network nodes of a brain area are constantly stimulated for a period of the characteristic time of the nodes, $\eta^{-1}$ to evoke damped oscillations with a maximum magnitude of one. The stimulation response of an isolated node is subtracted from the response of stimulated nodes in the network. A *Principal Component Analysis* (PCA) was performed using the covariance matrix among the 16,500 nodes. The period of 0.5 s data after 0.5 s of stimulus onset was decomposed. For further analysis, up to three components are considered that cover more than 99 % of variance across conditions.

*Subspace similarity, clustering and dissipation networks* The dot product of the normalized eigenvectors from the decomposition the stimulation response was used to measure the similarity of the dissipation across different stimulation sites for a range of values of the balance of the SCs and a spatial range of the homogeneous SC. The eigenspaces are clustered based on the similarity measure using k-means for each SC-balance and each range of the homogeneous SC. The number of clusters is estimated via the gap statistic (Tibshirani et al., 2001). For each cluster, the eigenspaces are rotated to the basis of the one with the highest similarity among all in the cluster, using the singular value decomposition and calculating the optimal rotation matrix (Kabsch, 1978). Averaging the aligned basis vectors in a cluster (across eigenspaces) gives the set of dissipation networks for each cluster.

*Statistics on dissipation networks*. A Kolmogorov-Smirnov test is performed to determine whether the cortical and the subcortical contributions are drawn from the same distribution. A Wilcoxon rank-sum test is used to determine whether the cortical and the subcortical contributions to a dissipation network are equivalent. A significance level of 0.01 is used for both of these tests.

*Comparing dissipation and resting state (RS) networks*. Guided by the Brodmann area designation of the Automated Anatomical Labeling Template (Tzourio-Mazoyer et al., 2002) the cartographic description of the RS-networks by Damoiseaux et al. (2006) is mapped onto the geometrical model of the cortex and its parcellation used here. In Damoiseaux et al. (2006), cortical structures are either mentioned or explicitly emphasized to be part of a RS-network, but not explicitly excluded. For the present purposes, we assumed areas that were not mentioned were also not part of a RS-network. Finally, in the time since their 2006 publication, there have been a number of updates to the functional designation of the different RS-networks. We have kept the original designations save for the 'unspecified' RS-network, which seems to best correspond the dorsal attention network (Cole et al., 2010).

The resultant map onto our geometrical model describes the probability of an area to contribute to a RS-network by three levels (no, medium, or high contribution). The Bhattacharyya coefficient (Bhattacharyya, 1946), *BC* is then used to estimate the amount of overlap between a RS- and a dissipation network, which is essentially an eigenvector. The square of each eigenvector element is taken and summed up within each area. The coarse-grained eigenvectors and each sum of a combination thereof (in total 4) are normalized to unit length. RS-networks and dissipation networks are compared using the *BC* for a RS-network and each normalized coarse-grained eigenvector or combination thereof. The significance of each comparison, $p = (n + 1) / (N + 1)$ is estimated by *N*-times permuting the entries of a RS-network (without replacement), calculating the coefficient, $\widetilde{BC}$ (the permuted Bhattacharyya coefficient) and then counting the values greater than the original, $n : \widetilde{BC}_i > BC$, with $N = 2 \times 10^6$. The *p*-values are corrected due to 24 independent multiple comparisons (eight RS-networks and three eigenvectors), using the Bonferroni-Holm-correction. A *BC* with *p*-values less than 0.05 is considered to be significant. The mean across the maximum significant overlap for the RS-networks with a dissipation network (i.e., a single eigenvector or a combination thereof) gives the optimal parameters for (i) the used eigenvector coarsening metric (i.e., absolute or squared value); (ii) the balance of the homogeneous and the heterogeneous SCs; and (iii) the spatial range of the homogeneous SC. The optimum parameter set is determined for dissipation networks due to cortical, subcortical and both cortical and subcortical stimulations.

*Comparing dissipation and connectivity structure*. A dissipation network gives the contribution of a brain area (mean contribution of nodes within) to a functional network. The spatial structure is specific to each of the dissipation networks that best explain an experimentally observed RS-network (from **Fig. 5**). These eight dissipation networks are compared to the heterogeneous SC. Because this SC describes the wiring between brain areas, the role of each brain area within the topology is characterized using measures from graph theory, namely: in-, out-, total-degree; in-, out-, total-strength; and clustering coefficient) (Rubinov and Sporns, 2010). Incoming, outgoing, or all connected ties to an area are measured in terms of (i) their numbers, and (ii) their weights. By counting the connections we obtain the in-, the out-, and the total-degree. By calculating the sum of connection weights we obtain the in-, the out-, and the total-strength. The clustering coefficient measures the degree to which areas in a graph tend to group together. Each of the seven topological characterizations of brain areas is then compared to the functional contribution to a dissipation network, using the *BC*. To test statistical significance, the same permutation test is used as for the comparison of the dissipation with the RS-networks.

## Results

Following stimulation of a cortical area, the induced activity initially spreads radially from the stimulation site across area boundaries, due to local and homogeneous structural connectivity (SC). Then, propagation occurs across large distances through the brain network via heterogeneous SC, that is, white matter tracts (**Fig. 3b**). In the SC-matrices, the connection weights and the time delays impose strong constraints on the propagation of activity. Thus, stimulation of adjacent areas may cause totally different propagation patterns (**Fig. 3b**). Conversely, stimulation at two remote sites may lead to similar spatiotemporal pattern after an initial transient (see, time frame 890 ms in **Fig. 3b**).

*Dissipation networks*. From our parametric study, we find a maximum of eleven different dissipation networks for a ratio of 20 % homogeneous SC to 80 % heterogeneous SC and a spatial range for homogeneous SC between 30 mm and 35 mm (**Fig. 4a**). With a pure heterogeneous SC, only four networks can be identified; while the number of dissipation networks decreases as the proportion of homogeneous SC increases (**Fig. 4a**). We find a maximum of 27 areas in two occurrences: for a 40 % / 60 % homogeneous/heterogeneous SC-ratio and a spatial range of 38 mm for the homogeneous SC, and in the case of 100 % heterogeneous SC (**Fig. 4b**). We conclude that although a pure heterogeneous SC can produce several dissipation networks, considering homogeneous SC dramatically increases the repertoire of activated networks. However, there is an optimal value, as too much homogenous SC is detrimental to the richness of the repertoire.

*Resting state (RS-)networks*. From the decomposition of the response activity we obtain a maximum of three dissipation networks per stimulation site capturing 99 % of the dissipation. We thus assessed (i) whether these dissipation networks correlate with the experimentally observed RS-networks (Damoiseaux et al., 2006), and (ii) whether the set of RS-networks is mainly driven by cortical, subcortical or both structures. Interestingly, the optimal ratio of heterogeneous/homogeneous SCs is found to be 20 % / 80 % consistently for all stimulation conditions. The optimal spatial range for the homogeneous SC is found to be 10 mm. The locations of the stimulation that are most likely to support dissipation into one of the RS-patterns are listed in **Table 2** (with its corresponding correlation (Bhattacharyya) coefficient) for each stimulation condition and for the optimal parameterization. Note that a location may appear repeatedly for the same stimulation condition, because the activity after stimulation is decomposed into the three dissipation networks, where each of which may activate a different RS-network (e.g., area AD in thalamus).

Irrespective of the restrictions to the stimulation, the default mode and the memory network always show the highest correspondence with the dissipation networks, whereas the visual and the auditory network show the lowest correspondence (**Table 2**). Moreover, we averaged the best coefficients (**Table 2**) over RS-networks to decide whether either cortical or subcortical areas drive the given set of eight RS-networks best, or eventually a mix of cortical and subcortical stimulations. Considering the overall correspondence, the set of RS-networks is equally well explained by stimulating subcortical sites ($<BC>$ = 0.77 on average) than cortical sites ($<BC>$ = 0.77), but best explained by stimulating a mixture of both, cortical and subcortical sites ($<BC>$ = 0.79). The dissipation networks matching best with the RS-networks are shown in **Fig. 5**.

To assess whether a dissipation network reflects the underlying structure, we correlated the graph metrics of areas in each the dissipation networks with the corresponding measures in the heterogeneous SC (**Figure 6**). Across the different measures, the in-degree of the SC can be related to the two memory networks and the attention network.

*Stimulation look up table.* The dissipation and RS-networks can be characterized in terms of stimulation sites. Thalamic stimulations result in activity most prominently in four RS-networks: default mode, motor, working memory and the attention network. Cortical stimulations, in particular superior temporal, primary motor, secondary visual, and anterior cingulate cortex result in activity most prominently in the remaining RS-networks, namely auditory-phonological, somato-motor, memory, and ventral stream network. All stimulation sites for cortical and subcortical areas of which their dissipation networks significantly match with a RS-network are summarized in **Figure 7**. Note that the cortical areas that trigger a particular dissipation network, especially memory, working memory and somato-motor are scattered over the cerebral hemispheres (**Fig. 7a**). In addition, **Figure 7** indicates which of the three dissipation networks matches with a RS-network. Considering that the order of dissipation networks reflects the captured variance in a particular response to a specific stimulation, we found the following RS-networks to be dominant in terms of variance, thus captured in the first dissipation network: the visual, the auditory, the motor and the working memory networks. The same is true, to a lesser extent, for the memory, the ventral stream and the attention network. These RS-networks were represented in the specific second dissipation network to stimulation, thus weaker in terms of the variance of the particular responses. Interestingly, we found the default mode network to be particularly flexible and spanned by both the first dissipation network and the second dissipation network due to specific stimulation.

**Discussion**

Here we show that resting state (RS-)networks can be specifically activated following the stimulation of specific brain areas. To enable such activation we hypothesized that networks operate at the brink of criticality. So far, predictions related to near-criticality have only been tested in autonomous situations of ongoing spontaneous brain activity (Deco et al., 2009, 2011, 2012; Ghosh et al., 2008; Hansen et al., 2015; Honey et al., 2007). In non-autonomous situations, such as following stimulations of individual brain areas, (near-)criticality, which is linked mathematically to the local center manifold theorem (Haken, 1978), predicts that the post-stimulus dynamics of dissipation networks evolve with characteristic features in space and time: (i) the existence of a low-dimensional set of dissipation networks, and (ii), their slow decay times after a stimulus relative to other networks. This approach provides not only a link between brain stimulations and RS-networks (as suggested by Fox et al., 2014), but also gives a better understanding of the relation between sensory external inputs and internal brain states. Along these lines, RS-dynamics originate from subspaces, in which the ongoing activity evolves and alters, giving rise to non-stationarity as observed in empirical and computational studies (Allen et al., 2014; Hansen et al., 2015). Our study predicts that these subspaces can be selectively targeted to bias the brain dynamics towards the activation of specific RS-networks through stimulation of specific brain areas, for instance, by sensory stimulation (e.g., auditory, visual) and brain stimulation techniques (e.g., transcranial magnetic stimulation). The stimulation sites are predicted to be network-specific and spatially clustered but distributed (**Fig. 7**). Stimulating different brain areas, for instance by *transcranial magnetic stimulation*, could lead to similar activation patterns during rest conditions.

The criticality of the brain network model essentially depends on (i) the distance of the node's operating point to the bifurcation, (ii) the effects of the SC on the nodes' operating point, (iii) the ensemble of signal transmission delays, and (iv) the stimulation. In the absence of stimulations the operating point of the network is always near-critical. Excitatory stimulations push the network model closer to criticality. We have demonstrated that the dissipation networks show near-criticality and represent the most active and long-lasting patterns following stimulation. Mohajerani et al. (2013) demonstrated in lightly anesthetized or awake adult mice that a palette of sensory-evoked and hemisphere-wide activity motifs is represented in spontaneous activity.

Correlation analysis between functional circuits and intracortical axonal projections indicated a common framework corresponding to long-range monosynaptic connections between cortical areas. Mohajerani et al. (2013) also report that most of the robust activation patterns and their evolution appeared long after stimulation, reflecting that the initial dynamics are determined by the local interactions and the stimulation site but the later developments are shaped by the interplay of connectome and dynamics. These results converge with our findings and suggest that an intracortical monosynaptic connectome shapes the spatiotemporal evolution of spontaneous cortical activity.

The SCs mostly predict the activity of brain areas directly after stimulation. However, as time evolves, both implemented types of SC, local (homogeneous) SC and large-scale (heterogeneous) SC, play a crucial role in the spatiotemporal progress. The connectome and its large-scale SC can explain some, but not all dissipation networks that fit the experimentally observed RS-networks best (**Fig. 6**). It is interesting to note that the default mode and the memory networks strongly depend upon the SC, which suggests that they play a special role in information processing. The activation of the other RS-networks depends to a lesser degree on the SC and thus constitutes an emergent dynamic process. Emergent properties can be understood by the transmission and synchronization behavior of the oscillatory activities throughout the propagation in the network, which decelerates or accelerates the dissipation process in parts of the network. The comparison with the SC (**Fig. 6**) indicates that the dissipation processes are oscillatory sequences over multiple cycles, where delays and synchronization naturally play a major role.

Our simulations show that the dynamic repertoire of dissipation networks is the richest for the mixed case in which local and large-scale SC are simultaneously present (**Fig. 4**), in keeping with known statistics of synapses within a population, namely 50 % of intracortical and 50 % of corticocortical fibers (Braitenberg and Schüz, 1998). We identified the 80 % / 20 % ratio and 10 mm values as optimal to obtain all RS-networks. These values were similar for all areas, but it is likely that they are brain-area specific. However, we did not perform an area-specific optimization, as the number of possibilities makes it computationally intractable at the current time. Not only could the synaptic connections be better adapted to predict the empirical data, but there are also possibilities for improving the characteristics of the local dynamics in each brain area. At the moment the regional local dynamics are considered homogeneous as a matter of simplification, but could be extended to deal with different heterogeneous local dynamical nodes, for instance, derived from the temporal information in functional data (Deco and Kringelbach, 2014b).

In conclusion, we demonstrated that connectome based large-scale networks dissipate their energy spatiotemporally upon stimulation in a characteristic low-dimensional manner, which is consistent with the idea that ongoing RS-dynamics operate close to criticality. The dissipation networks are compatible with the empirically known RS-networks and are set apart by the slow time scale as predicted by theorems of near-criticality. Stimulation sites can be assembled in topological groups that approximate empirical RS-networks. A stimulation of brain areas in these groups predicts an evolution of the RS-dynamics towards lower-dimensional subspaces, in which the subsequent dynamics evolve and can be characterized by conventional functional connectivity (FC) approaches. Our results suggest a means to bias RS-dynamics via spatially coordinated stimulation towards target subspaces. Given that FC of the RS differentiates groups with different pathologies and across ages, our results are of interest for approaches of such brain stimulation techniques as *transcranial alternating current stimulation* (tACS), *transcranial magnetic stimulation* (TMS), and *deep brain stimulation* (DBS) directed towards therapy and cognitive enhancement.

# Figures and tables

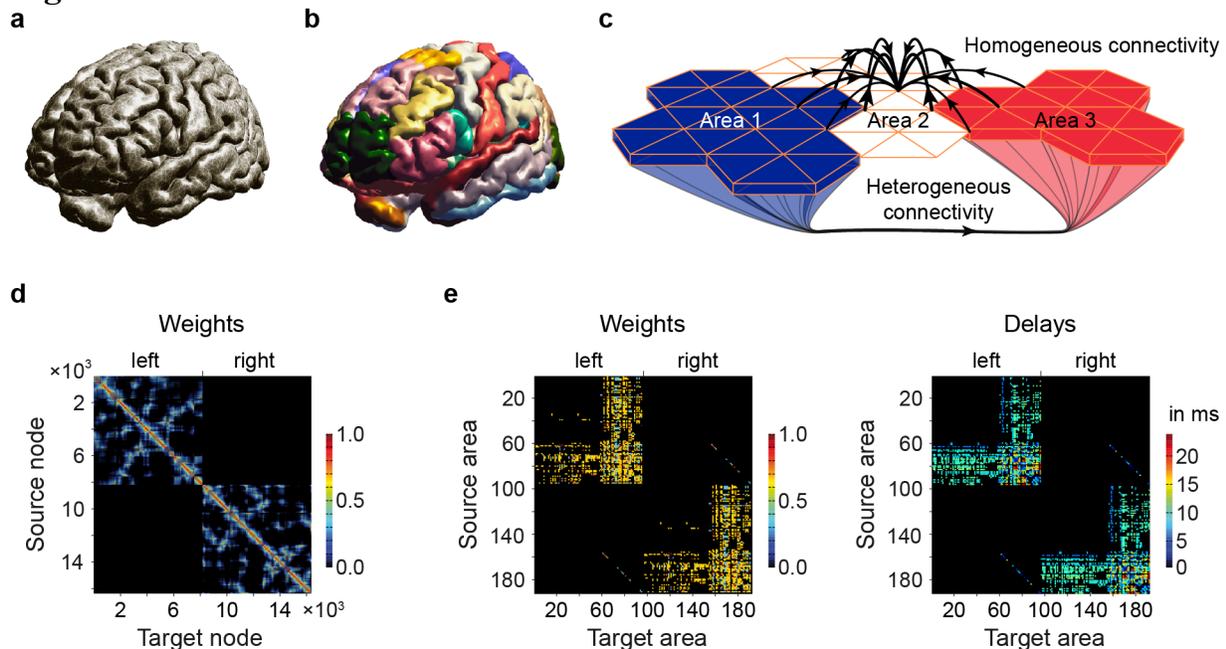

**Figure 1.** Structure of the large-scale brain model. The large-scale brain model is composed of (a) the brain's geometry of 116 subcortical areas and the two cerebral hemispheres. There are 37 cortical areas (b), each containing between 29 and 683 nodes (dots in (a)), for a total of 8,192 nodes per hemisphere. (c) Homogenous and heterogeneous structural connectivity (SC). Heterogeneous SC corresponds to white matter tracts connecting brain areas over long distances. Homogenous SC corresponds to gray matter fibers, with local connections within a given area, but also enabling some communication over short distances between neighboring areas. Although Area 2 is not connected to Areas 1 and 3 via the white matter, it is weakly linked to both areas via a set of local SC. (d) Homogeneous SC matrix for the 16,384 nodes. The synaptic weights are color coded. The diagonal describes in warm colors the strong SC of adjacent nodes. SC decreases with distance, which is shown in cold colors. SC of nearby nodes are scattered (e.g., blue dots) in (d) because each cerebral hemisphere is described by a surface, which makes it impossible to cluster nodes locally along both axes. Note the absence on interhemispheric local SC. (e) Heterogeneous SC for the 190 (74 cortical + 116 subcortical) areas for weights (left panel) and time delays (right panel). Within one hemisphere, the 58 subcortical areas mostly project to the 37 cortical ones. Some connections between subcortical areas can also be seen. The 37 cortical areas project heavily to both cortical and subcortical areas. Some interhemispheric connections can also been seen. Note also the presence of large time delays.

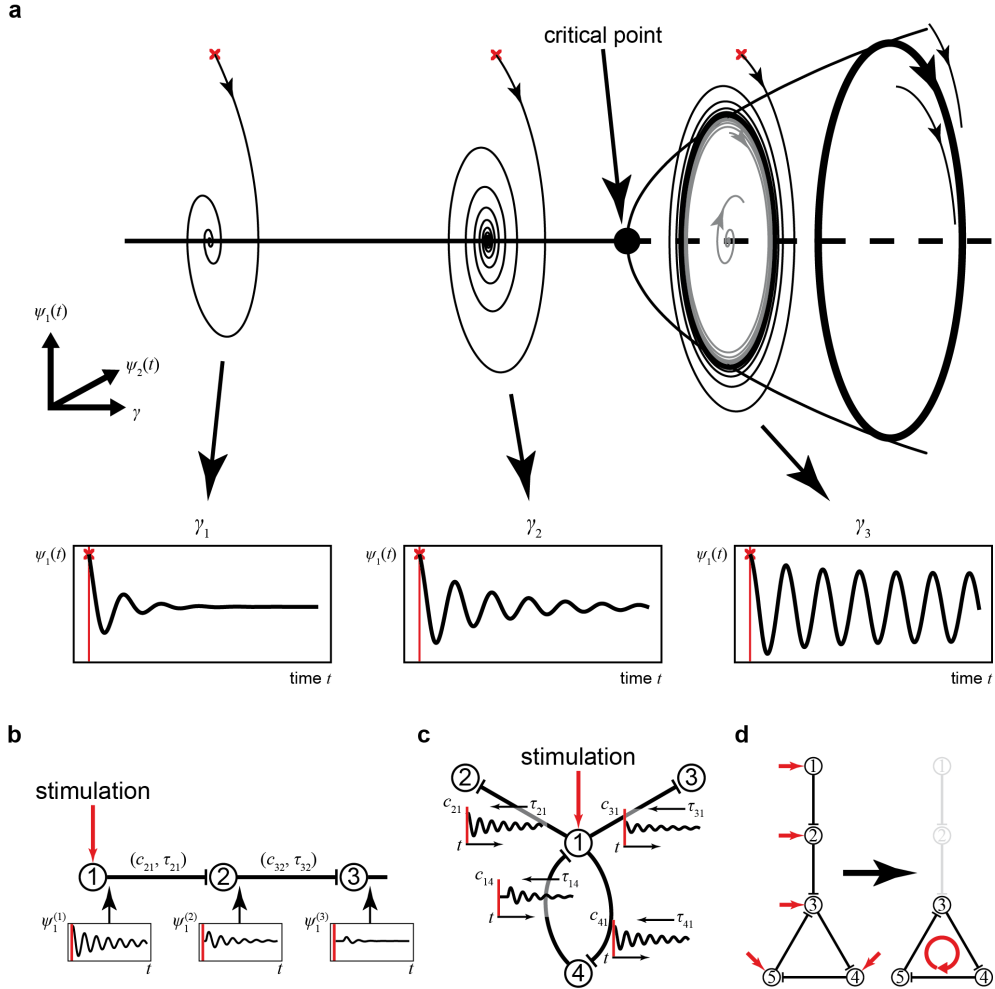

**Figure 2.** The large-scale brain model works near criticality. (**a**) Each node in the model is parameterized by $\gamma$ to operate at the same distance from the critical point. A node shows zero activity or oscillation (~42 Hz) in response to stimulation (red crosses). The activity at each node is described by two time-dependent variables, $\psi_1(t)$ and $\psi_2(t)$. The closer a node operates to the critical point, the larger and the longer lasting is the oscillation (compare $\gamma_1$ and, $\gamma_2$). When the critical point is reached, the node intrinsically performs a rhythm of constant magnitude. The model, however, is set so that the critical point is never exceeded. (**b**) Principles of activity spreading after stimulation. The damped oscillation generated in the stimulated node (1) is sent via its efferent connections to its target node (2), triggering there, in turn, a damped oscillation with weaker amplitude and faster decay, which then propagates to the next node. Activity $\psi_1^{(j)}(t)$ of node ($j$) is scaled by $c_{ij}$ and transmitted to node ($i$) via homogeneous and heterogeneous connections (SCs), delayed by $\tau_{ij}$ in the latter case. In such a chain, activity would decay fast. (**c**) In the large-scale brain model, multiple activity re-entry points can be found. At any time point, the dynamics of a node is influenced by all incoming activity. The node's response to stimulation (1) is relayed to linked nodes (2-4), which may be fed back to (1) via (4) and may allow the induced activity to dissipate on a much longer time scale. The network response thus depends upon the SCs and allows the network to operate near criticality. (**d**) Activation of dissipation networks. Activity after stimulating a node (1 or 2) in a series connection decays fast (as in **b**). However, activity may circulate and thus decays slower in a feedback network (4-5). Such remaining activity after the initial stimulation decay reveals the so-called dissipation networks.

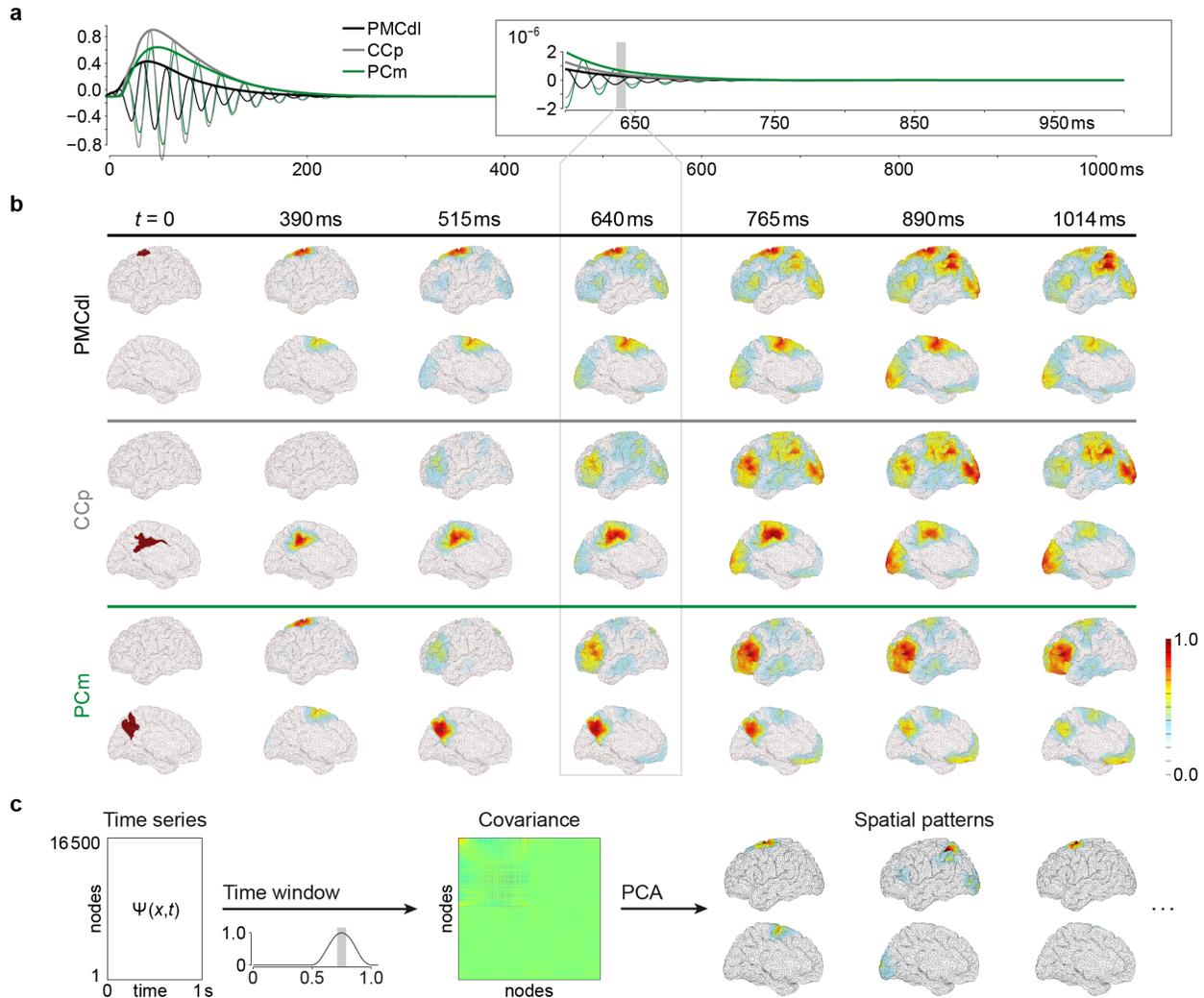

**Figure 3.** Dissipation after stimulation. (**a**) Response of area PFCcl to the activation of three different regions PMCdl, CCp and PCm (abbreviations are given in **Table 1**). Note that the amplitude, decay and phase of the response depend upon the stimulated area. The main determinants of the response pattern are the connections, the synaptic weights and the time delays. The envelope of the time series is computed (black, gray and green lines for the three stimulation sites). (**b**) Spatiotemporal activation following stimulation of three different regions. At a given time point, we extract the amplitude of the envelope for the 16,500 nodes (the 16,384 cortical nodes and the 116 subcortical ones), which we normalize to 1. The color scale thus indicates the contribution of a given region to the overall activity. The dissipation of activity after stimulating two distant brain areas, PMCdl and CCp (located far from one another: PMCdl in the lateral surface, CCp in the medial surface) leads to similar topographical patterns (for $t > 640$ ms). In contrast, a distinct pattern appears when stimulating PCm, which is adjacent to CCp. (**c**) Extraction of the main activated propagation subnetworks. We use the stimulation of PMCdl as an example. We calculate the covariance among the 16,500 time series (the 16,384 cortical nodes and the 116 subcortical ones) for a time window centered at 750 ms and then perform a Principal Component Analysis (PCA) to extract the subnetworks capturing >99 % of the activity. Three different subnetworks are thus activated when PMCdl is stimulated.

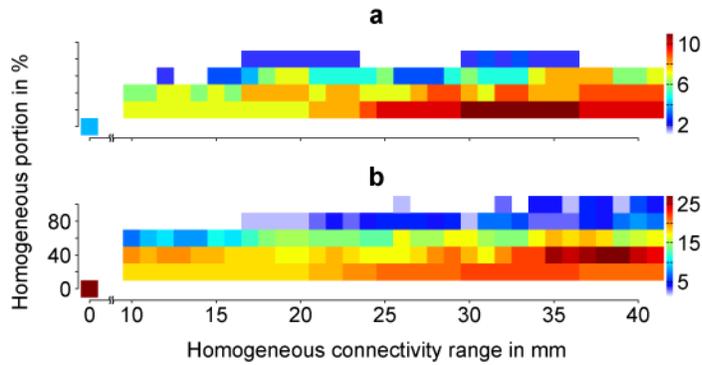

**Figure 4** Repertoire of dissipation networks. The set of dissipation networks (**a**), and the number of effective stimulation sites (**b**) depend on the spatial range of the homogeneous SC and the ratio of homogeneous to heterogeneous SCs.

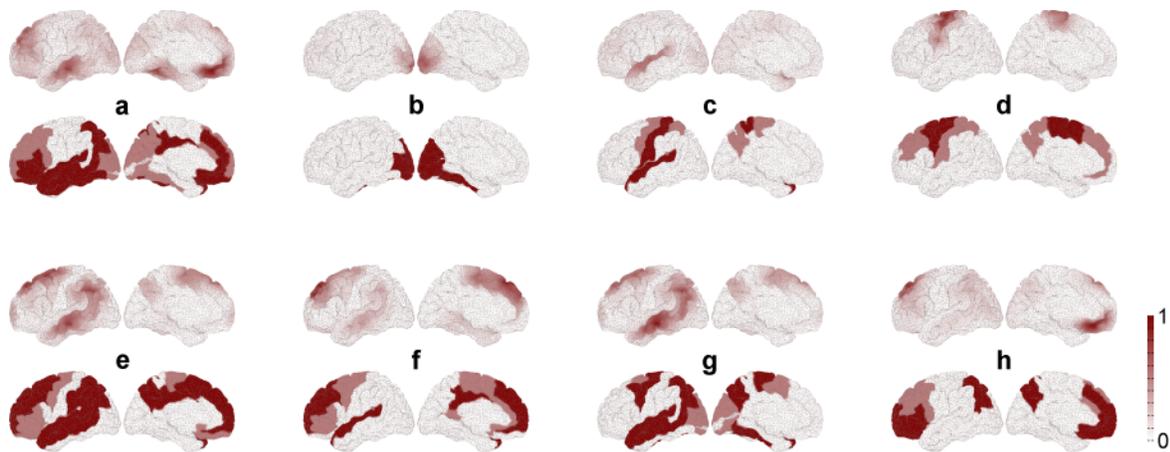

**Figure 5** Comparison between dissipation networks (top rows) and the experimentally observed RS-networks (bottom rows) for the lateral and medial surface of the brain. From **a**–**h**: default mode, visual, auditory-phonological, somato-motor, memory, ventral-stream, dorsal attention and working memory. We used 20 % / 80 % for the ratio of heterogeneous/homogeneous SCs and a range of 10 mm for the homogeneous SC. The white to red scale gives the relative contribution of areas to the dissipation (top rows) and the RS-networks (bottom rows). The stimulation sites are given in **Table 2** and **Fig. 7**.

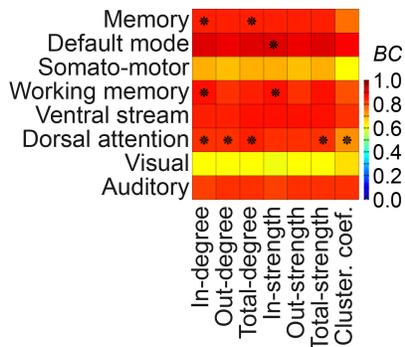

**Figure 6** Influence of the structure on the RS-like networks. Each dissipation network (from **Fig. 5**) that best explains an experimentally observed RS-network (rows) is correlated with the underlying heterogeneous SC using seven graph-theoretic measures (columns). Incoming, outgoing, or all connected ties to an area can be measured in terms of number, i.e., in-, out-, total-

degree, or in terms of strength, i.e., in-, out-, total-strength. The clustering coefficient measures the degree to which areas in a graph tend to cluster together. *BC* indicates a matching with warmer colors, where comparisons marked with a star are statistically significant. Note that correlations may be high but not significant using a permutation test. The in-degree of the heterogeneous SC can be related to the two memory networks and the attention network. The activation of the other RS-networks is emergent properties, which cannot be predicted by the architecture.

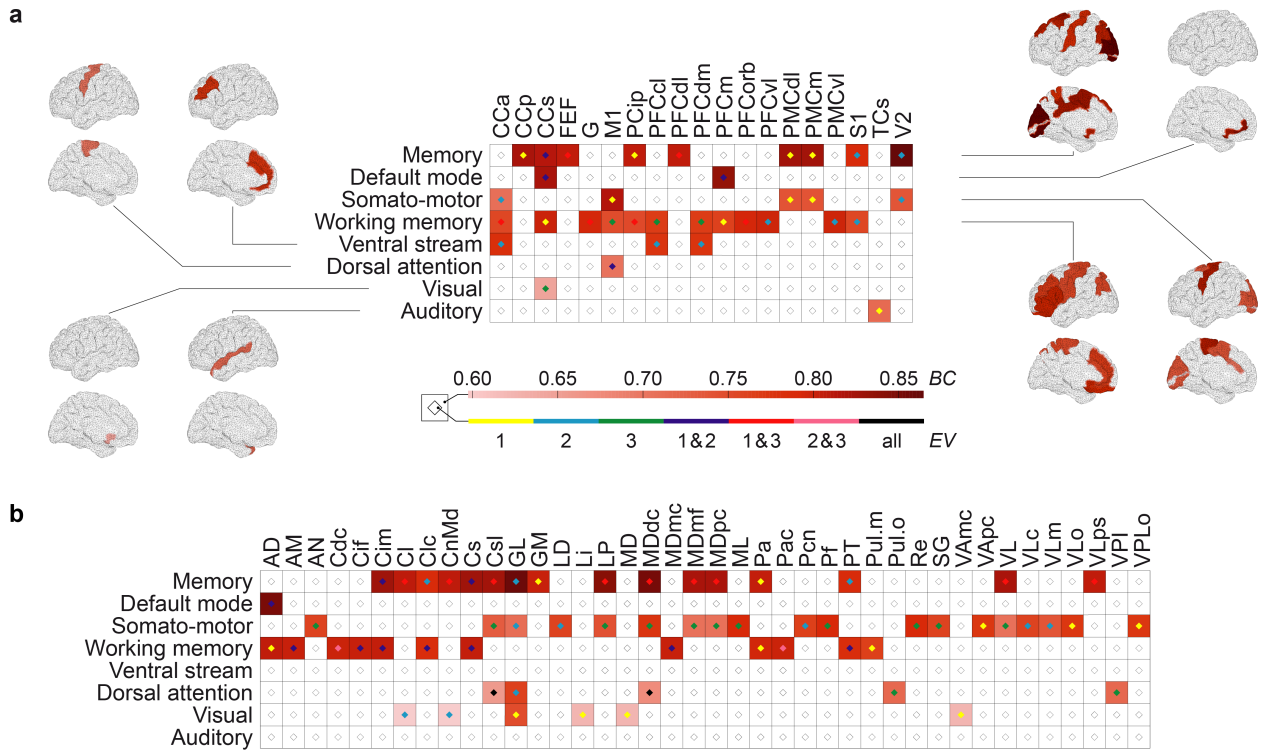

**Figure 7** RS-like networks triggered by stimulation. Cortical stimulations in **a,** and subcortical in **b** lead to dissipation networks correlating significantly with RS-networks for a ratio of 20 % / 80 % of the heterogeneous/homogeneous SCs and a range of 10 mm of the homogeneous SC. *BC* = [0, 1] indicates a matching with higher values. The eigenvectors, *EV* (1 to 3 in descending order of eigenvalues), indicate the dissipation network(s) matching with RS-networks. Abbreviations are listed in **Table 1**. Note that the sites triggering a particular pattern can be scattered over the cerebral hemispheres (e.g., for the two memory networks and the somato-motor network).

**Table 1** Abbreviations of brain areas. Number of nodes per cortical areas in brackets (left,right).

| | | | |
|---|---|---|---|
| A1 | Primary auditory cortex (57,74) | Cld | Capsule of the nucleus lateralis dorsalis |
| A2 | Secondary auditory cortex (33,64) | CnMd | Nucleus centrum medianum thalami |
| Amyg | Amygdala (151,135) | Cs | Nucleus centralis superior thalami |
| CCa | Gyrus cinguli anterior (54,49) | Csl | Nucleus centralis superior lateralis thalami |
| CCp | Gyrus cinguli posterior (167,179) | GL | Nucleus geniculatus lateralis thalami |
| CCr | Gyrus cinguli retrosplenialis (68,67) | GM | Nucleus geniculatus medialis thalami |
| CCs | Gyrus cinguli subgenualis (29,42) | GMpc | Nucleus geniculatus medialis thalami, pars parvocellularis |
| FEF | Frontal eye field (104,161) | IL | Intralaminar nuclei of the thalamus |
| G | Gustatory cortex (52,42) | LD | Laterodorsal nucleus (thalamus) |
| HC | Hippocampal cortex (75,54) | Li | Nucleus limitans thalami |
| Ia | Anterior insula (48,71) | LP | Nucleus lateralis posterior thalami |
| Ip | Posterior insula (82,111) | MD | Nucleus medialis dorsalis thalami |
| M1 | Primary motor area (463,460) | MDdc | Nucleus medialis dorsalis thalami, pars densocellularis |
| PCi | Inferior parietal cortex (454,371) | MDmc | Nucleus medialis dorsalis thalami, pars magnocellularis |
| PCip | Cortex of the intraparietal sulcus (355,486) | MDmf | Nucleus medialis dorsalis thalami, pars multiformis |
| PCm | Medial parietal cortex (196,241) | MDpc | Nucleus medialis dorsalis thalami, pars parvocellularis |
| PCs | Superior parietal cortex (199,177) | ML | Midline nuclei of the thalamus |
| PFCcl | Centrolateral prefrontal cortex (328,227) | Pa | Nucleus paraventricularis thalami |
| PFCdl | Dorsolateral prefrontal cortex (248,216) | Pac | Nucleus paraventricularis caudalis thalami |
| PFCdm | Dorsomedial prefrontal cortex (211,270) | Pcn | Nucleus paracentralis thalami |
| PFCm | Medial prefrontal cortex (61,68) | Pf | Nucleus parafascicularis thalami |
| PFCorb | Orbital prefrontal cortex (310,265) | PT | Nucleus parataenialis thalami |
| PFCpol | Pole of prefrontal cortex (279,279) | Pul | Nucleus pulvinaris thalami |
| PFCvl | Ventrolateral prefrontal cortex (380,479) | Pul.i | Nucleus pulvinaris inferior thalami |
| PHC | Parahippocampal cortex (267,212) | lPul.l | Nucleus pulvinaris lateralis thalami |
| PMCdl | Dorsolateral premotor cortex (108,138) | Pul.m | Nucleus pulvinaris medialis thalami |
| PMCm | Medial premotor cortex (149,68) | Pul.o | Nucleus pulvinaris oralis thalami |
| PMCvl | Ventrolateral premotor cortex (126,138) | R | Nucleus reticularis thalami |
| S1 | Primary somatosensory cortex (487,420) | Re | Nucleus reuniens thalami |
| S2 | Secondary somatosensory cortex (107,116) | SG | Nucleus suprageniculatus thalami |
| TCc | Central temporal cortex (436,422) | Teg.a | Nucleus tegmentalis anterior |
| TCi | Inferior temporal cortex (390,306) | VA | ventral anterior nucleus (thalamus) |
| TCpol | Pole of temporal cortex (91,101) | VAmc | Nucleus ventralis anterior thalami, pars magnocellularis |
| TCs | Superior temporal cortex (306,352) | VApc | Nucleus ventralis anterior thalami, pars parvocellularis |
| TCv | Ventral temporal cortex (260,317) | VL | ventral lateral nucleus (thalamus) |
| V1 | Visual area 1 (147,180) | VLc | Nucleus ventralis lateralis thalami, pars caudalis |
| V2 | Secondary visual cortex (683,663) | VLm | Nucleus ventralis lateralis thalami, pars medialis |
| AD | Nucleus anterior dorsalis thalami | VLo | Nucleus ventralis lateralis thalami, pars oralis |
| AM | Nucleus anterior medialis thalami | VLps | Nucleus ventralis lateralis thalami, pars postrema |
| AN | Anterior nuclei of the thalamus | VP | Nucleus ventralis posterior |
| AV | Nucleus anterior ventralis thalami | VPI | Nucleus ventralis posterior inferior thalami |
| Caud | Nucleus caudatus | VPL | Aentral posterior lateral nucleus (thalamus) |
| Cdc | Nucleus centralis densocellularis thalami | VPLc | Nucleus ventralis posterior lateralis thalami, pars caudalis |
| Cif | Nucleus centralis inferior thalami | VPLo | Nucleus ventralis posterior lateralis thalami, pars oralis |
| Cim | Nucleus centralis intermedialis thalami | VPM | Nucleus ventralis posterior medialis thalami |
| Cl | Nucleus centralis lateralis thalami | VPMpc | Nucleus ventralis posterior medialis, pars parvocellularis |
| Clau | Claustrum | X | Area X (thalamus) |
| Clc | Nucleus centralis latocellularis thalami | | |

**Table 2** Stimulation site(s) of the dissipation network that match a RS-network best. Each given experimentally known RS-network is compared to all dissipation networks, which is triggered by stimulation for a parameterization. Stimulation is restricted to either cortical or subcortical sites, otherwise all sites are considered. For each stimulation condition, the optima parameterization is found for the best accordance of dissipation networks and the whole set of RS-networks. The optimal parameterization is a ratio of 20 % / 80 % for the heterogeneous/homogeneous SCs and a range of 10 mm for the homogeneous SC, simply the range is with 17 mm different on the condition of subcortical stimulations. Note the mix of cortical and subcortical sites (last column) that drives best the set of RS-networks. The coefficient in parenthesis is the matching values (it varies between 0 and 1). Abbreviations are listed in **Table 1**.

| Resting state network | Stimulation condition | | |
| --- | --- | --- | --- |
| | Cortex (excl. subcortex) | Subcortex (excl. cortex) | Cortex and Subcortex |
| Default mode | PFCm (0.8337) | AD (0.8420) | AD (0.8506) |
| Visual | CCs (0.6455) | GL (0.6953) | GL (0.7510) |
| Auditory-phonological | TCs (0.7147) | GMPC (0.6630) | TCs (0.7147) |
| Somato-motor | M1 (0.8153) | MDDC (0.8199) | M1 (0.8153) |
| Memory | V2 (0.8646) | MDDC (0.8454) | V2 (0.8646) |
| Ventral stream | CCa (0.7845) | ML, AN, SG (0.8122) | CCa (0.7845) |
| Dorsal attention | M1 (0.7039) | R, VA, X (0.7097) | AD (0.7631) |
| Working memory | CCs (0.8006) | PAC, Cdc (0.8204) | GL (0.8069) |